\documentclass[12pt,prc,nopacs,nokeys,nodate,superscriptaddress,nofootinbib]{revtex4}
\usepackage{amsfonts}
\usepackage{amsbsy}
\usepackage{mathrsfs}
\usepackage{graphicx,epsfig,rotating}
\usepackage{color,pstricks}

\newcommand{\turnt}[1]{\begin{turn}{90}#1\end{turn}}

\begin{document}
\title{Exploring hybrid star matter at NICA and FAIR}
\author{T.~Kl\"ahn}
\affiliation{Institute for Theoretical Physics, University of Wroc{\l}aw,
Poland}
\author{D.~Blaschke }\thanks{e-mail: david@theor.jinr.ru}
\affiliation{Institute for Theoretical Physics, University of Wroc{\l}aw,
Poland}
\affiliation{Joint Institute for Nuclear Research,
Dubna, Russia}
\author{F.~Weber}
\affiliation{Department of Physics, San Diego State University, USA}
\begin{abstract}
  We discuss constraints for the equation of state of hybrid star
  matter which can be obtained from next generation 
  heavy-ion collisions at FAIR and NICA.  
  Particular emphasis is on the planned NICA facility 
  at JINR Dubna which shall provide fixed-target and collider experiments 
  just in the relevant energy ranges.
\end{abstract}
\maketitle

Neutron stars (NS) provide valuable insights into the nature of
nuclear matter at densities several times beyond nuclear saturation
($n_s\approx0.16$ fm$^{-3}$)
\cite{Weber:1999qn,Glendenning:2000,Klahn:2006ir}.  In particular any
extreme (either small or large) value of measured neutron star
observables, such as radius, mass, and temperature, is likely to
improve our understanding of the properties of cold and dense matter
substantially.  A prime example underlining this statement is neutron
star PSR J1614-2230, whose recently measured mass of
$(1.97\pm0.04)$~M$_\odot$ makes this object the
heaviest neutron star ever observed with sufficient accuracy and
confidence \cite{Demorest:2010bx}.  This measurement has direct
consequences for the investiagtion of dense matter in terrestrial
heavy-ion collision (HIC) experiments, as planned for NICA 
\cite{Sorin:2011zz} and FAIR \cite{Staszel:2010zz}.  
The reason for this is the close relationship that exists between the
stiffness of the equation of state (EoS) of symmetric nuclear matter
and the highest NS mass supported by the associated EoS for neutron
star matter, as illustrated in Figs.~1 and 2 in \cite{Klahn:2006ir}.
In this paper, a testing scheme, consisting of different constraints
from astrophysical observations and HIC experiments, has been
introduced and consistently applied to a given set of nuclear EoS.
In detail, the scheme suggests that a viable EoS should
\begin{itemize}
\item reproduce the most massive observed neutron star,
\item avoid the direct URCA (DU) cooling problem,
\item result in neutron stars within the predicted mass-radius domains
  of 4U 0614+09 (deduced from quasiperiodic brightness
  oscillations) and RX J1856-3754 (deduced from the objects thermal emission),
\item explain the gravitational mass and total baryon number of
pulsar PSR J0737-3039(B) with at most 1\% deviation from the baryon
number predicted for this particular object,
\item not contradict flow and kaon production data of heavy-ion
collisions~.
\end{itemize}

At the time of publication of Ref.\ \cite{Klahn:2006ir}, the most massive NS 
has been PSR J0751+1807 with $M \sim 2.1$~M$_\odot$, a result which has later 
been withdrawn \cite{Nice:2007}. 
A second, less strict mass
constraint had been formulated, which demanded that the successful
nuclear EoS reproduces at least a maximum neutron star mass of $1.6$
M$_\odot$.  
All the EoS investigated back then passed this constraint.
Due to the precision of the new mass measurement for PSR J1614-2230,
we now update the overall result of this testing scheme (analogously
to Table V in \cite{Klahn:2006ir}).  
It is evident from Table \ref{tab:sum} that only four (five) of the 
formerly eight nuclear EoS are still compatible with the upper (lower) 
mass value measured for PSR J1614-2230. 
An additional model for the nuclear EOS--the hybrid EOS ``DBHF-NJL''--
has been included in Table \ref{tab:sum}, which has several advantages 
over the purely hadronic EoS. 
Details about this EOS will be discussed below.
  
In summary, the important lesson that we learn from high NS masses, as
measured for PSR J1614-2230, is that the EoS has to be 'rather stiff'
or, conversely, can `not be too soft' at ultra-high densities in order
to be compatible with neutron star masses.
In the following we point out how HIC experiments 
can contribute to constrain the behavior of the nuclear EoS further
and, most importantly, provide most valuable information about the possible
existence of quark matter in compact stars.

\begin{table}[!tbh]
\begin{center}
\begin{tabular}{|l||cc|cc|cc|cc|cc|cc||cc|}
\hline
{Model}                                 &
\turnt{$M_{\rm max}\ge 2.01~M_\odot$}            &
\turnt{$M_{\rm max}\ge 1.93~M_\odot$}            &
\turnt{$M_{DU}\ge 1.5~M_\odot$}         &
\turnt{$M_{DU}\ge 1.35~M_\odot$}        &
\turnt{4U 1636-536 (u)}& 
\turnt{4U 1636-536 (l)}& 
\turnt{RX J1856 (A)} & 
\turnt{RX J1856 (B)}                            &
\turnt{J0737 (no loss) }                &
\turnt{J0737 (loss 1\% $M_\odot$)}              &
\turnt{SIS+AGS flow}                    &
\turnt{SIS flow+ K$^+$ prod.}                         &
\turnt{No. of passed tests}                 &
\turnt{(out of 6)}      \\
\hline
NL$\rho$ 	  	&$-$  &$-$  &$-$&$-$  &$-$  &$-$&$-$  &$-$  &$-$  &$-$  &$+$  &$+$ &$1$&$1$\\[-2mm]
NL$\rho\delta$ 	&$-$  &$-$ &$-$&$-$  &$-$ &$-$&$-$ &$-$ &$-$ &$-$ &$+$&$+$ &$1$&$1$\\[-2mm]
{\bf DBHF}     		&$+$  &$+$  &$-$&$-$  &$+$ &$+$ &$-$  &$+$  &$-$  &$+$  &$-$&$+$ &$2$&$5$\\[-2mm]
{\bf DD}       		&$+$  &$+$  &$+$ &$+$  &$+$ &$+$&$-$  &$+$  &$-$  &$-$  &$-$ &$-$  &$3$&$4$\\[-2mm]
{\bf D$^3$C}   		&$+$  &$+$  &$+$ &$+$  &$+$ &$+$ &$-$  &$+$  &$-$  &$-$  &$-$ &$-$  &$3$&$4$\\[-2mm]
KVR      		&$-$  &$-$ &$+$ &$+$  &$-$ &$\circ$ &$-$  &$-$  &$-$  &$+$  &$+$ &$+$ &$2$&$4$\\[-2mm]
{\bf KVOR}     		&$+$  &$+$  &$+$ &$+$   &$-$ &$+$&$-$  &$-$  &$-$  &$\circ$&$+$&$+$&$3$&$5$\\[-2mm]
DD-F     		&$-$  &$+$  &$+$ &$+$   &$-$ &$+$&$-$  &$-$  &$-$  &$+$  &$+$&$+$ &$2$&$5$\\
\hline
{\bf DBHF+NJL}          &$+$ & $+$  &$*$ &$*$  &$+$  &$+$&$-$&$+$&$-$&$+$&$+$&$+$ &$4$&$6$\\
\hline
\end{tabular}
\end{center}
\caption{\label{tab:sum}
  Summary of results for the testing scheme suggested in \cite{Klahn:2006ir},
  updated by the recently measured mass of PSR J1614-2230
  ($(1.97\pm0.04)$ M$_\odot$) and supplemented by  the hybrid EoS of 
  Ref.~\cite{Klahn:2006iw} (DBHF-NJL, last line), for which results are shown 
  in Fig.~1.
  EoS not labeled in bold face fail to reproduce a NS mass of $2.01~M_\odot$.
  {Non-separated columns show the results for a strict (left) and weakened 
    (right) interpretation of the corresponding constraint.}
  {The last column gives the total number of tests passed by a given EOS.}
  (For details, see \cite{Klahn:2006ir}.)
}
\end{table}

We focus here on two topics in HIC experiments which are relevant for
determining the stiffness of nuclear matter at supersaturation
densities and thus have direct implications for the astrophysics of
compact stars: (i) strangeness production and 
(ii) transverse and elliptical particle flow.

\begin{figure}[!htb]
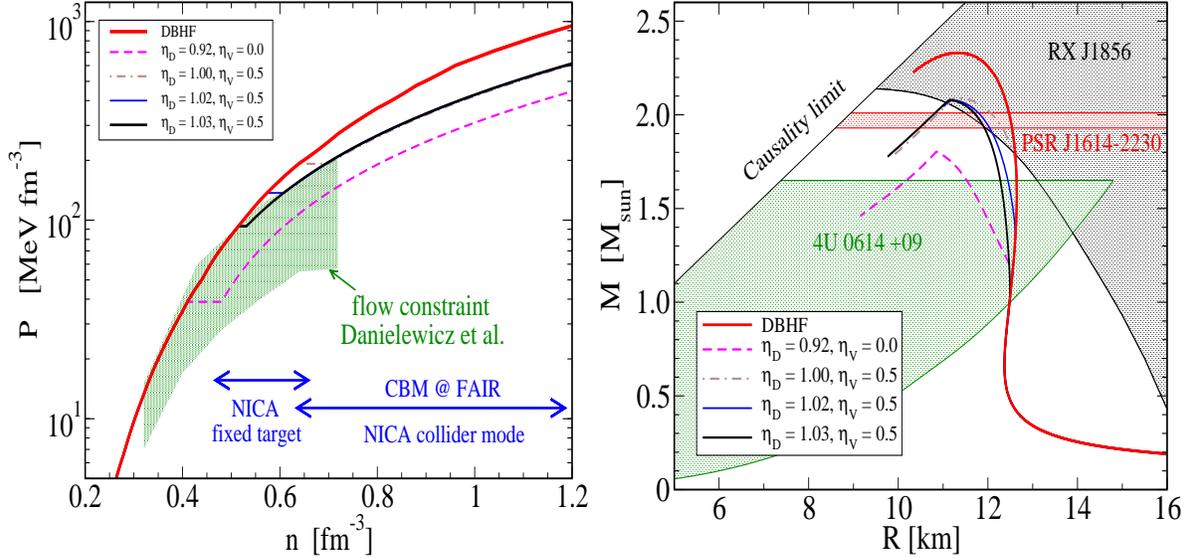

\begin{tabular}{cc}
\includegraphics[height=0.45\textwidth,width=0.47\textwidth,angle=0]{Flow-nica.eps}
&
\includegraphics[height=0.45\textwidth,width=0.47\textwidth,angle=0]{M-R-nica.eps}
\end{tabular}
\caption{Left panel: Flow constraint \cite{Danielewicz:2002pu}
  extracted from HIC experiments in the range $E_{\rm lab}=0.4$ to
  10~A~GeV and estimated regions accessible at CBM and NICA experiments
  \cite{Arsene:2006vf,Blaschke:2010ka}. Right panel: Sequences of
  compact star M-R relations obtained from solution of the
  Tolman-Oppenheimer-Volkoff \cite{Weber:1999qn,Glendenning:2000} 
  equation for different EoS without (DBHF) and with deconfinement transition, 
  as compared to  NS constraints.  }
\label{fig:eosconstraint}
\end{figure}

According to the present status of the theory, subthreshold $K^+$
production at $E_{\rm lab}<1.58$~A~GeV appears to require a
sufficiently soft equation of state (see \cite{Fuchs:2005zg} and 
\cite{Hartnack:2011cn} for recent reviews). 
This is an interesting complement 
to the astrophysical requirement of sufficiently stiff equations of state 
demanded by the observations of high-mass neutron stars, as discussed above.
The analysis of flow data for experiments at different energies
($E_{\rm lab}=0.4$ to 10~A~GeV in Ref.~\cite{Danielewicz:2002pu}) put
a constraint on the cold symmetric nuclear EoS, which represents a
region in the nuclear pressure-density plane, $P(n)$, shown in
Fig.~\ref{fig:eosconstraint}.
This constraint is readily applied to the isospin-symmetric part of
any neutron star EoS and can therefore be used to derive upper bounds
on the maximum mass of compact stars \cite{Klahn:2006ir}, as can be
seen by comparing the left and right panels of
Fig.~\ref{fig:eosconstraint}.
From the latter constraint it is evident
that an independent confirmation of this flow constraint is very important,
both theoretically as well as experimentally by providing sufficiently accurate 
flow data in the region of $E_{\rm lab}\sim 1$ to 5~A~GeV 
(Nuclotron fixed-target range), where we suspect the deconfinement phase 
transition to occur, and just above the limits of the AGS data range, i.e. for 
$E_{\rm lab}= 10$ to 60~A~GeV (CBM range and fixed-target equivalent of the 
NICA collider range).  
According to the flow constraint, the {\it ab-initio} DBHF
EoS, computed for the Bonn-A nucleon-nucleon potential,
appears to be too stiff at densities above $3.5$ times saturation
density. Assuming the deconfinement phase transition to
  provide the softening of the high-density EoS just in this density
  range and at the maximally tolerable stiffness of the 
  Danielewicz {\it et al.} constraint would result in an upper limit for 
  the maximum mass of compact stars at $2.1$ M$_\odot$
\cite{Klahn:2006ir,Klahn:2006iw}.
This is compatible with the new mass constraint provided by PSR J1614-2230.
The quark matter EoS used here is the three-flavor color
superconducting NJL model with self-consistently determined masses and
diquark gaps \cite{Blaschke:2005uj}, including a vector meson
meanfield which entails a sufficient stiffening of the hybrid EoS
\cite{Klahn:2006iw} (a nonlocal, covariant generalization has been
provided in Ref.~\cite{Blaschke:2007ri}).  
Neglecting the latter term would result in a too soft hybrid EoS, 
conflicting with the new neutron star mass constraint (dashed lines in
Fig~\ref{fig:eosconstraint}).  Moreover, if the diquark
pairing interaction would be sufficiently reduced or even be neglected
altogether so that no color superconducting phase can occur, the phase
transition would occur at densities too high to be realized even within 
the most massive NS.

Due to isospin asymmetry, in compact stars the deconfinement
transition always occurs at lower baryon densities than in symmetric
nuclear matter.  
There is another, indirect argument to expect a deconfinement phase 
transition in symmetric matter at densities not exceeding about 3 to 4 
nuclear saturation densities.  
This relatively low critical density would correspond to a deconfinement 
transition before the onset of hyperon formation.
This is a possible solution to the problem that hyperon EoS are often too
soft to fulfill the maximum-mass constraints, see Ref.~\cite{Baldo:2003vx}.
In addition, a sufficiently low onset density for quark matter avoids the DU 
problem of the DBHF EoS which otherwise would lead to a too fast cooling of 
neutron stars with masses in the typical binary radio pulsar range 
$1.3\le M/M_\odot \le 1.4$.

As valuable as the analysis of NS observables and the
  evaluation of kaon production and flow data appears, presently
  applied model EoS for hybrid matter suffer from the artificial
  description of the hadron-to-quark matter phase transition.  
  Two-phase approaches, using Maxwell or Glendenning constructions
  \cite{glen91:pt}, cannot make trustworthy statements
  about the phase transition region, and only estimate the region of a
  transition very roughly.  In general, these constructions describe
  only first order transitions.  Therefore, any statement about the
  critical endpoint in the phase diagram which results from
  corresponding models is, at best, an estimate.  
  More elaborate treatments applying functional renormalization group 
  methods are promising  \cite{Schaefer:2007pw}
  but have to be extended to include the baryons.
  Even then they will be lacking to descibe aspects resulting from the 
  compositeness of baryons such as the existence of a scattering continuum 
  of three- and multiquark states at finite densities which entails the 
  dissociation of baryons into their quark constituents \cite{Wang:2010iu}.
  A generalized Beth-Uhlenbeck EoS \cite{Schmidt:1990,Hufner:1994ma}
  which accounts for these effects is presently being developed 
  \cite{Zablocki:2011}.

In conclusion, the question whether quark matter exists in compact stars is 
very challenging \cite{Alford:2006vz} and closely related to the question of
deconfinement in heavy-ion collisions at CBM \cite{Blaschke:2008cu} 
and NICA \cite{Blaschke:2010ka}.  
As we don't know a priori the coupling strengths in various interaction
channels at high densities, it is important to test the conjectured 
EoS in a heavy-ion collision experiment.  
The energies provided at the NICA facility
(both fixed target and collider) are perfectly suited
for providing astrophysically relevant constraints. 
Without further necessary experimental evidence as will be provided, e.g., 
from NICA or FAIR one cannot safely rule out or support the possibility that 
all neutron stars in the observed mass range between 1.23 and 2.01~$M_\odot$ 
are hybrid stars with
quark matter cores!

\subsection*{Acknowledgement}
D.B. is grateful for the hospitality 
{granted}
to him during a visit at San Diego 
State University and acknowledges support from ``CompStar'', 
a research networking programme of the European Science foundation, 
from a grant of the Polish Ministry of Science and Higher Education 
supporting this programme and from the Russian Fund for Basic Research 
under grant No. 11-02-01538-a.
The work of T.K. was supported in part by ``hadronphysics2'' within the 
European framework programme FP7.
F.W. is supported by the National Science Foundation (USA)
under Grant PHY-0854699.

\end{document}